\documentclass{cmspaper}
\usepackage{graphicx}
\usepackage{epsfig} 
\begin{document}
%
\begin{titlepage}
\title{Monte Carlo study of gg$\rightarrow$H+jets contribution to Vector Boson Fusion Higgs production at the LHC}
\begin{Authlist}
A. Nikitenko\Instfoot{cern}{Imperial College, London; on leave from ITEP, Moscow.}
M. V\'azquez Acosta\Instfoot{cern}{CERN, European Organization for Nuclear Research, Geneva.}
\end{Authlist}

\begin{abstract}

The contribution of $gg \rightarrow H+jets$ production process to the vector boson
fusion production of the Higgs boson, $VV \rightarrow H$, was evaluated
with the ALPGEN generator and the PYTHIA shower Monte Carlo including a jet-parton matching 
procedure. After the experimental like event selections applied at PYTHIA 
particle level, the contribution was found to be 4-5 \% for a Higgs boson mass of 
120 GeV.  

\end{abstract}
\end{titlepage}
\section{Introduction}

The cross section measurements of the Higgs boson production in the vector boson fusion
(VBF) process at LHC, $VV \rightarrow H$ ($qq \rightarrow qqH$), followed by Higgs boson 
decays into $\tau \tau$, $WW$ and $\gamma \gamma$ will significantly extend the 
possibility of Higgs boson coupling measurements~\cite{coupl1,coupl2}. 
According to the latest full simulation CMS results \cite{cmsphy} the most promising
VBF channel in the Higgs boson mass range of 115-135 GeV is
$qq \rightarrow qqH, ~ H \rightarrow \tau \tau$ \cite{tautau}. For the higher
Higgs boson mass the best VBF channels in CMS are 
$H \rightarrow W W ^{\ast} \rightarrow \ell \ell \nu \nu$ \cite{ww_ll} and   
$H \rightarrow W W ^{\ast} \rightarrow \ell \nu \rm j \rm j$ \cite{ww_ljj}.

The uncertainty of the coupling measurement using VBF channels will depend on 
the contribution of $gg \rightarrow H+jets$ process after 
event selections. The parton level, leading order calculations 
\cite{delduca1, delduca2, delduca3} have shown that the fraction of selected events 
due to this process can be as large as 30\% after VBF selections for a Higgs boson mass of 120 GeV. 
The effect of QCD corrections to $gg \rightarrow H+\rm j \rm j$ process in the 
Higgs boson mass region of 115-160 GeV was found to be 
15-26 \% before VBF selections and 30-40 \% after $\eta$ separation between
two highest $p_{T}$ jets was applied \cite{John}. 


We present a new estimate of the contribution of the $gg \rightarrow H+jets$ process 
using the ALPGEN~\cite{alpgen} generator with the MLM prescription for jet-parton matching 
\cite{Mangano:2006rw, Hoche:2006ph} at the PYTHIA shower simulation \cite{pythia} in the
case in which the Higgs boson mass is 120 GeV.
  
\section{Event generation and simulation}

The VBF Higgs boson production was generated with the PYTHIA version 6.409.
The leading order (LO) cross section of $qq \rightarrow qqH$ process given 
by PYTHIA is 4.22 pb. The $gg \rightarrow H+jets$ production was generated using ALPGEN 
version 2.06 with the MLM prescription for jet-parton matching. The parton shower simulation was 
performed using PYTHIA 6.409. The CTEQ5L PDF was used in both ALPGEN and PYTHIA as well as 
the default values of the factorization and renormalization scales.

The parton level cuts applied in the ALPGEN generation are $p_t^{j}>20$ GeV, $|\eta^{j}|<5$ and $\Delta r_{jj}>0.5$. 
In the case of the H + n jets (n $\geq$ 2) generation, "soft" VBF phase space preselections at the parton level 
were applied:
\begin{itemize}
\item $M_{j1j2}>600$ GeV
\item $|\Delta\eta^{j1j2}|>4$,
\end{itemize}
where $M_{j1j2}$ is the invariant mass and $\Delta\eta^{j1j2}$ is the difference in pseudorapidity of the two 
leading $p_{T}$ partons. The parameters for MLM jet-parton matching were: $E_{T}^{clus}$=20 GeV, $R^{clus}$=0.5 
and $\eta ^{cl~max}=$5.0.

The jets at particle level, after showering and hadronization in PYTHIA, were found with the simple cone algorithm 
implemented in PYTHIA routine PYCELL. The parameters of the PYCELL jet finder are the following: the cone size is 0.5, 
the seed threshold is 2 GeV, the pseudorapidity coverage is 5.0 and the cell size in $\Delta \eta \times \Delta \phi$ is
$\sim$ 0.1 $\times$ 0.1 (granularity of the CMS hadron calorimeter).

For the PYTHIA underlying event model the Tune DWT \cite{tuneDWT} was used and the stability of the 
results were checked with the Tune A \cite{tuneA}. The PYTHIA parameters for both Tunes are listed 
in Table ~\ref{table1}.

\begin{table}[b]
\caption{Underlying Event Tunes used in PYTHIA}
\label{table1}
\begin{center}
\begin{tabular}[t]{|c|c|c|c|}\hline
 Parameter & Tune A & Tune DWT \\
\hline
MSTP(81) & 1 & 1 \\
MSTP(82) & 4 & 4 \\
PARP(82) & 2.0 GeV & 1.9409 GeV \\ 
PARP(83) & 0.5 & 0.5 \\ 
PARP(84) & 0.4 & 0.4 \\ 
PARP(85) & 0.9 & 1.0 \\ 
PARP(86) & 0.95 & 1.0 \\ 
PARP(89) & 1.8 TeV & 1.96 TeV\\ 
PARP(90) & 0.25 & 0.16\\ 
PARP(62) & 1.0 & 1.25\\ 
PARP(64) & 1.0 & 0.2 \\ 
PARP(67) & 4.0 & 2.5 \\ 
\hline
\end{tabular}
\end{center}
\end{table}

The number of ALPGEN generated $gg \rightarrow H+jets$ events and cross sections given by ALPGEN
are shown in Table~\ref{table2}.
\begin{table}
\caption{The number of ALPGEN generated $gg \rightarrow H+jets$ events and cross sections given by ALPGEN.}
\label{table2}
\begin{center}
\begin{tabular}[t]{|c|c|c|c|}\hline
 Sample & N generated events & VBF preselection & $\sigma$ (pb) \\
\hline
H + 1 jet & 329196 & No & 19.54 \\
H + 2 jets & 26825 & Yes & 0.693 \\
H + 3 jets & 5513 & Yes & 0.574 \\
H + 4 jets & 1326 & Yes & 0.355 \\
\hline
\end{tabular}
\end{center}
\end{table}

The ALPGEN generated events were passed through the MLM jet-parton matching procedure to avoid double counting. 
Table ~\ref{table3} shows the number of selected events for a given matching type, matching
efficiency and cross sections after matching. 
\begin{table}
\caption{The MLM matching type, number of selected events, matching efficiency and cross-sections.}
\label{table3}
\begin{center}
\begin{tabular}[t]{|c|c|c|c|c|c|}\hline
 Sample & matching type & N selected events & matching efficiency & $\sigma$ (fb) \\
\hline
H + 1 jet & exclusive & 100000 & 0.30  & 5936  \\
H + 2 jets & exclusive & 2307 & 0.09 & 59.7   \\
H + 3 jets & exclusive & 333 & 0.06 & 34.7   \\
H + 4 jets & inclusive & 224 & 0.17 & 60.1   \\
\hline
\end{tabular}
\end{center}
\end{table}

\section{Event selection}

The final VBF selections used in a full simulation analysis \cite{tautau} were applied to the PYTHIA particle 
level jets. An event must have at least two leading $E_{T}$ jets that satisfy the following requirements:
\begin{itemize}
\label{selection1}
\item $E_{T}^{j}>30$ GeV
\item $\eta^{j}~<~4.5$
\item $M_{j1j2}>1000$ GeV
\item $|\Delta\eta^{j1j2}|>4.5$
\item $\eta^{j1} \times \eta^{j2}<0$.
\end{itemize}
where j1 and j2 are two leading $E_{T}$ jets ordered in $E_{T}$.

The effect of applying a central jet veto was studied. The central jet veto requires to reject events
with a third jet that satisfies
\begin{itemize}
\item $E_{T}^{j3}>30$ GeV
\item $\eta^{j~min} + 0.5 < \eta^{j3} < \eta^{j~max} - 0.5$,
\end{itemize}
where $\eta^{j~min}$ and $\eta^{j~max}$ are the minimum and maximum $\eta$ of the two leading jets (j1 and j2).

\section{Results}

\begin{figure}[htb!]
\begin{center}
\epsfig{file=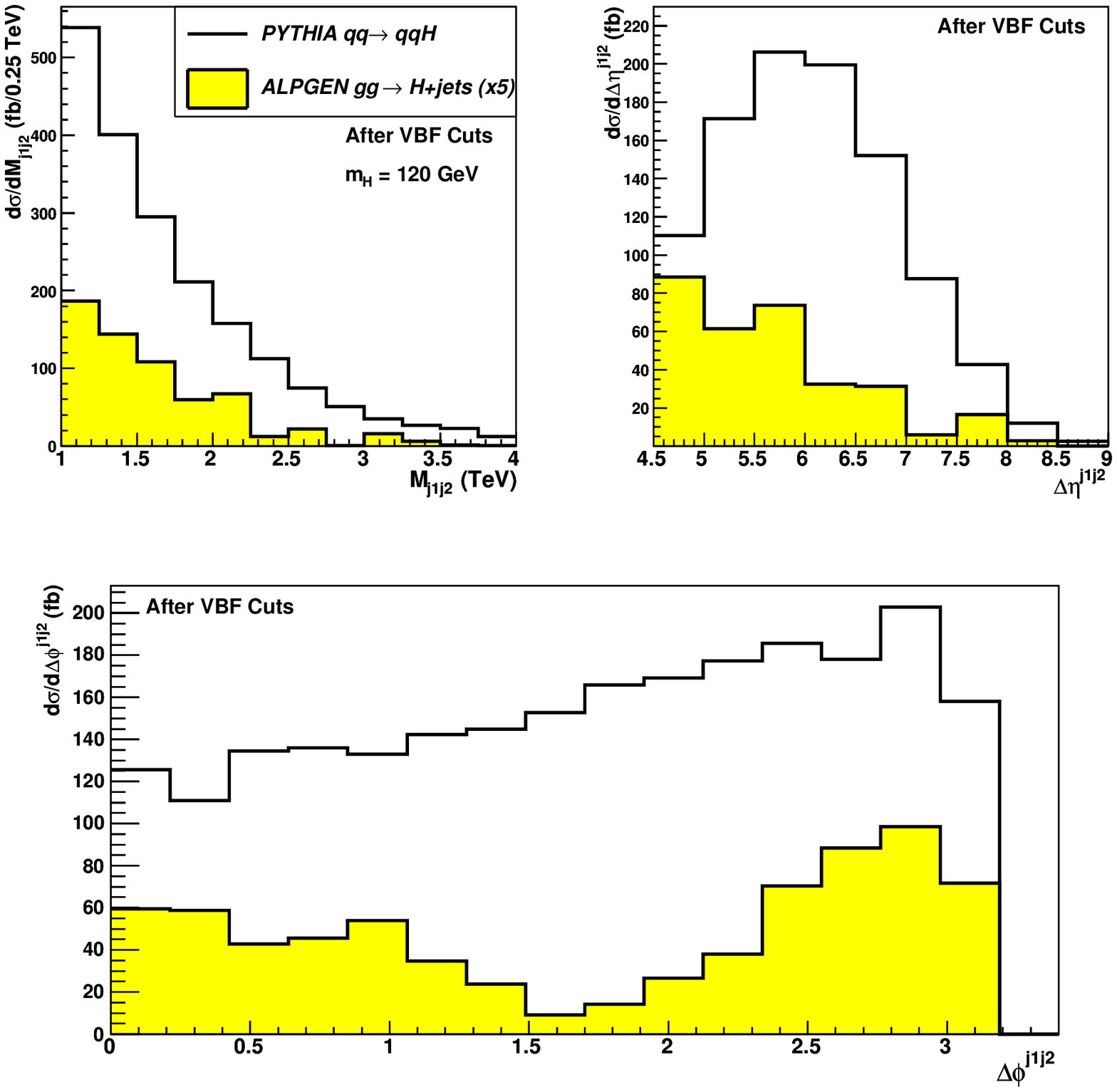,scale=0.9}
\caption{The differential cross section as a function of $M_{j1j2}$ (upper left plot), $|\Delta\eta^{j1j2}|$ 
(upper right plot) and $\Delta\phi^{j1j2}$ (bottom plot) for $qq \rightarrow qqH$ process (solid histogram) 
and $gg \rightarrow H+jets$ process (shaded histogram) after VBF selections. The cross sections for $gg \rightarrow H+jets$ 
process are shown multiplied by a factor 5.}
\label{Fig1}
\end{center}
\end{figure}

The cross section after VBF selections for $qq \rightarrow qqH$ is 492.3 fb and
for $gg \rightarrow H+jets$ is 31.3 fb, thus the contamination of $gg \rightarrow H+jets$ 
events after VBF selections is $\sim$ 6\%. The differential cross sections after VBF selections as a function 
of $M_{j1j2}$, $|\Delta\eta^{j1j2}|$ and $\Delta\phi^{j1j2}$ (azimuthal angle between the two jets in the transverse plane) 
are shown in Figure~\ref{Fig1} for both processes. The differential cross sections for $gg \rightarrow H+jets$ process 
shown in Figure~\ref{Fig1} are multiplied by a factor 5. The $\Delta\phi^{j1j2}$ distribution reflects the tensor 
structure of the couplings to vector bosons or gluons and can be used as a probe CP property of the couplings 
as proposed in \cite{cp1}, \cite{cp2}, \cite{delduca4}. The azimutal correlations between the two
jets in $gg \rightarrow H+\rm j \rm j$ process were found unchanged at NLO \cite{John}.
 
\begin{figure}[htb!]
\begin{center}
\epsfig{file=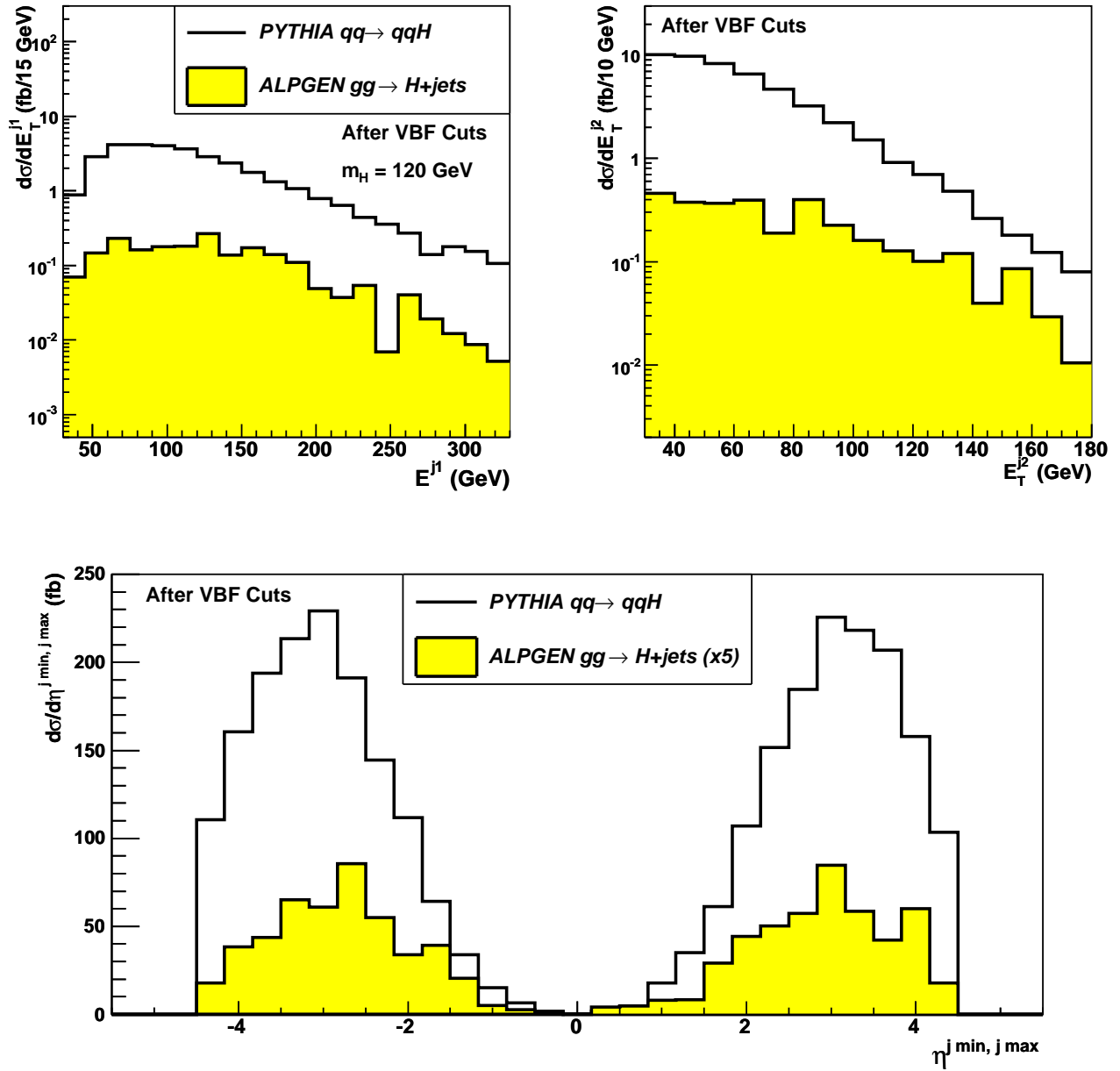,scale=0.9}
\caption{The $E_{T}$ and $\eta$ distributions of the two leading $E_{T}$ jets (j1 and j2) for 
$qq \rightarrow qqH$ process (solid histogram) and $gg \rightarrow H+jets$ process (shaded histogram) 
after VBF selections. The bottom plot shows $\eta$ distributions of the j1 and j2 with minimal ($\eta ^{j~min}$)
and maximal ($\eta ^{j~max}$) pseudorapidity, where the cross section for $gg \rightarrow H+jets$ 
process is shown multiplied by a factor 5.}
\label{Fig2}
\end{center}
\end{figure}

The $E_{T}$ and $\eta$ distributions of the two leading jets 
(j1 and j2) after VBF selections are presented in Figure~\ref{Fig2} normalized by the cross sections. 
The $\eta$ distribution for $gg \rightarrow H+jets$ process is shown multiplied by a factor 5.

\begin{figure}[htb!]
\begin{center}
\epsfig{file=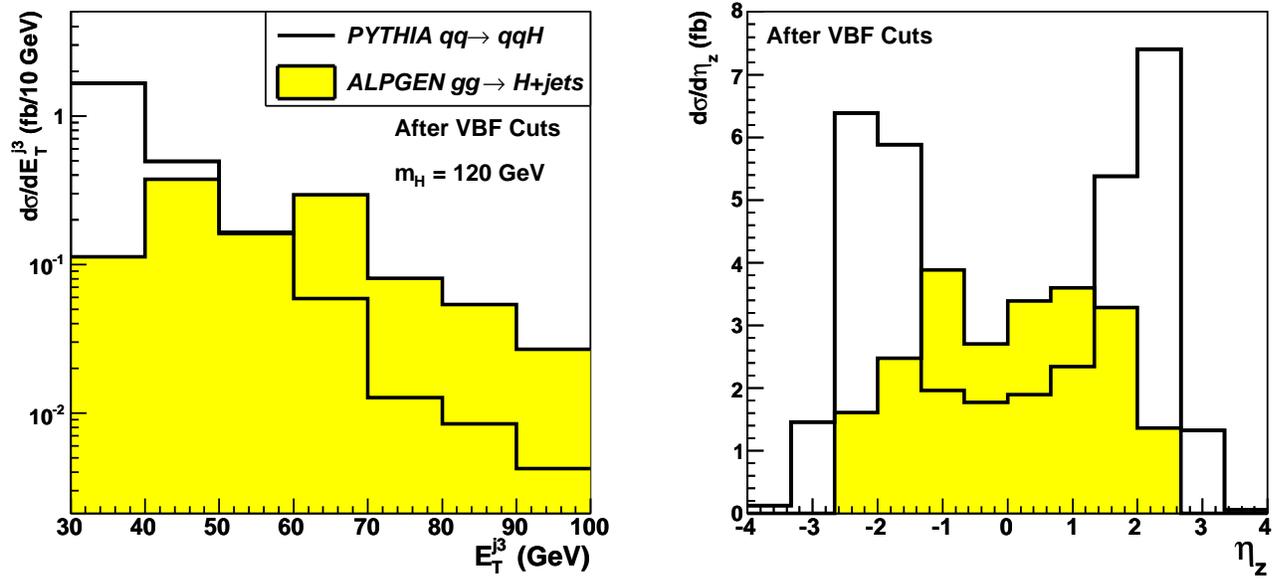,scale=0.9}
\caption{The $E_{T}$ and $\eta _{Z}$ distribution of the third jet (j3) for $qq \rightarrow qqH$ process 
(solid histogram) and $gg \rightarrow H+jets$ process (shaded histogram) after VBF selections. 
The $\eta _{Z}$ is defined as $\eta_{Z} = \eta^{j3} - 0.5 (\eta^{j1}+\eta^{j2})$.}
\label{Fig3}
\end{center}
\end{figure}

One of the key features of VBF Higgs boson production is, the so-called rapidity gap, due to an absence of the
color exchange in the t-channel \cite{rapgap1}, \cite{rapgap2}, \cite{rapgap3}. It leads to the lack of
the jet activity in the central detector region in contrast to the background processes to VBF Higgs boson:
$t \bar{t}$, QCD Z+jets, QCD WW+jets. The central jet veto was proposed as a tool to suppress background
both for heavy \cite{cjv1} and light \cite{cjv2} Higgs boson searches. The efficiency of
the central jet veto for $gg \rightarrow H+jets$ events was evaluated.
Figure~\ref{Fig3} shows the $E_{T}$ and $\eta _{Z}$ distribution of the third, highest $E_{T}$ jet in the
event with $E_{T}^{j3}>30$ GeV and in the pseudo-rapidity interval 
$\eta^{j~min} + 0.5 < \eta^{j3} < \eta^{j~max} - 0.5$ after VBF selections. The $\eta _{Z}$ is defined as 
$\eta_{Z} = \eta^{j3} - 0.5 (\eta^{j1}+\eta^{j2})$.

\begin{figure}[htb!]
\begin{center}
\epsfig{file=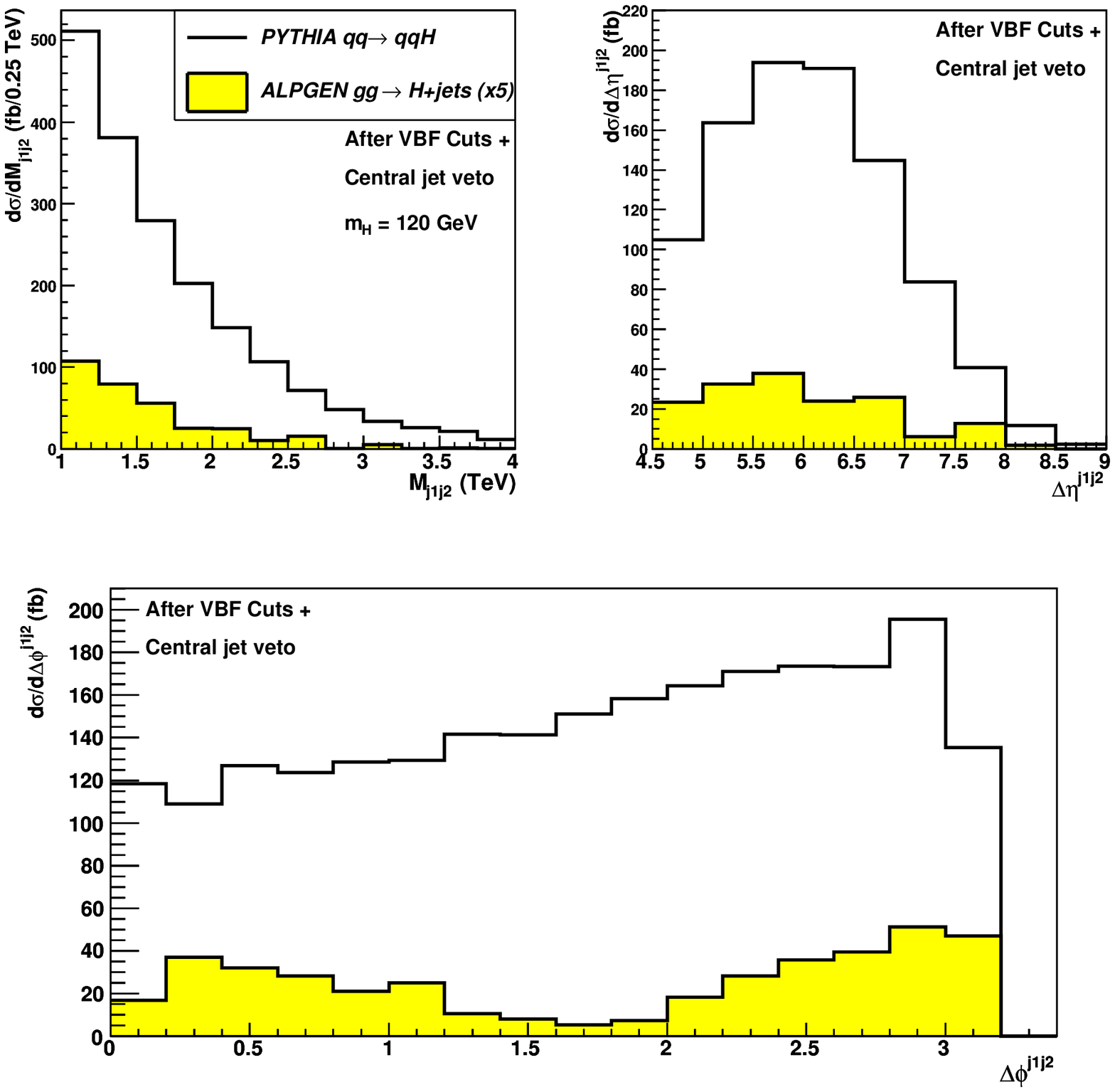,scale=0.9}
\caption{The same as Figure \ref{Fig1}, but after VBF and central jet veto selections.}
\label{Fig4}
\end{center}
\end{figure}

The total cross sections after VBF selections and the central jet veto is 468.3 fb for $qq \rightarrow qqH$
and 16.4 fb for $gg \rightarrow H+jets$, thus efficiency of the central jet veto for the "signal" VBF
events is 0.95 and for the "background" ($gg \rightarrow H+jets$) events is 0.52. After the central jet
veto, the contamination of $gg \rightarrow H+jets$ events is reduced from 6\% (after VBF selections) to
4 \%.  The fraction of the $gg \rightarrow H+1~jet$ cross section to the total 
$gg \rightarrow H+jets$ cross section is found to be only $\sim$ 1\% (0.24 fb). The differential cross sections after 
VBF selections and central jet veto as a function of $M_{j1j2}$, $|\Delta\eta^{j1j2}|$ and $\Delta\phi^{j1j2}$ 
are shown in Figure~\ref{Fig4}. The differential cross sections for $gg \rightarrow H+jets$ process 
shown in Figure~\ref{Fig4} are multiplied by a factor 5.

\subsection{Stability check of the ALPGEN results}

The $gg \rightarrow H+jets$ cross sections after VBF selections and central jet veto reported in the
previous section were obtained using the H+1jet to H+4jet ALPGEN samples with "soft" VBF preselections 
at generation (parton) level. As a cross check the cross sections were also evaluated using the H+1jet to 
H+3jet ALPGEN samples. Using the H+1jet to H+3jet samples only the cross section increased 
from 31.3 fb to 39.8 fb after VBF selections and from 16.4 fb to 20.0 fb after VBF selections plus 
central jet veto. 

For the case where H+1jet to H+3jet samples were used, the cross sections obtained with 
"soft" VBF preselections at parton level and with no preselections were compared. 
With no preselections the cross section increased from 20.0 fb to 24.6 fb after 
final VBF selections and central jet veto. The results are summarized in Table~\ref{table4}.
\begin{table}
\caption{Stability check of the results obtained with ALPGEN}
\label{table4}
\begin{center}
\begin{tabular}[t]{|c|c|c|c|}\hline
 VBF preselection & Samples & $\sigma(ggH)$ (fb) & $\sigma(ggH)/\sigma(qqH)$ \\
\hline
\multicolumn{4}{c}{After VBF selection} \\
\hline
Yes & H+1jet(exc)+H+2jet(exc)+H+3jet(exc)+H+4jet(inc) & 31.3 & 0.06 \\
Yes & H+1jet(exc)+H+2jet(exc)+H+3jet(inc) & 39.8 & 0.08 \\
\hline
\multicolumn{4}{c}{After VBF selection and Central jet veto} \\
\hline
Yes & H+1jet(exc)+H+2jet(exc)+H+3jet(exc)+H+4jet(inc) & 16.4 & 0.04 \\
Yes & H+1jet(exc)+H+2jet(exc)+H+3jet(inc) & 20.0 & 0.04 \\
No & H+1jet(exc)+H+2jet(exc)+H+3jet(inc) & 24.6 & 0.05 \\
\hline
\end{tabular}
\end{center}
\end{table}

Finally, results were re-evaluated with PYTHIA Tune A \cite{tuneA} and differences at the level less that 1 \%
with Tune DWT \cite{tuneDWT} were found.


\section{Conclusion}

The contribution of $gg \rightarrow H+jets$ events to the $gg \rightarrow qqH$ events after VBF selections and
central jet veto was estimated to be $\sim$ 4-5 \% for a Higgs boson mass of 120 GeV.
The result is stable within $\sim$ 25 \% to the usage or not of the "soft" VBF preselection 
and within $\sim$ 20 \% when ALPGEN samples are generated up to 3 or 4 jets. 
No effect on the results was found when PYTHIA Tune DWT or Tune A were used. 

\section{Acknowledgments}

We would like to thank M.~L.~Mangano, M.~Moretti and F.~Piccinini for the assistance in the generation of H+jets ALPGEN 
samples and for the useful discussions. We would also like to thank D.~Zeppenfeld for very useful discussions in the 
interpretation of the results obtained in this study.

\newpage

\newpage

\newpage

\newpage


\newpage

\end{document}